\documentclass{PoS}
\usepackage{amssymb, graphicx}

\usepackage[sort&compress]{natbib}

\bibpunct{[}{]}{,}{n}{}{}
\newcommand{\comment}[1]{}

\newcommand{\fm}{\, {\rm fm}}
\newcommand{\mev}{\, {\rm MeV}}

\newcommand{\lr}[1]{ \left( #1 \right) }
\newcommand{\lrs}[1]{ \left[ #1 \right] }

\renewcommand{\Im}{ {\rm Im} \, }

\newcommand{\sign}{ {\rm sign} \, }

\title{Deconfinement transition in two-flavour lattice QCD with dynamical overlap fermions.}

\ShortTitle{Deconfinement with dynamical overlap fermions.}

\author{\speaker{O.~A.~Kochetkov}\\
Institut f\"{u}r Theoretische Physik, Universit\"{a}t Regensburg, D-93040 Regensburg, Germany\\
Institute of Theoretical and Experimental Physics, 117259 Moscow, Russia\\
E-mail: \email{Oleg.Kochetkov@physik.uni-regensburg.de}
}

\author{V.~G.~Bornyakov
\thanks{Computations were performed on the ``Lomonosov''  supercomputer at the supercomputing center of MSU. V.B. is supported by the RFBR grant 13-02-01387-a.}\\
Institute for High Energy Physics, 142281 Protvino, Russia\\
Institute of Theoretical and Experimental Physics, 117259 Moscow, Russia\\
School of Biomedicine,Far Eastern Federal University, 690950 Vladivostok, Russia}

\author{P.~V.~Buividovich
\thanks{The work of P.B. was supported by the S.~Kowalewskaja award from the Alexander von Humboldt foundation.}\\
Institut f\"{u}r Theoretische Physik, Universit\"{a}t Regensburg, D-93040 Regensburg, Germany}

\author{N.~Cundy
\thanks{NC thank ``BK21 Plus Frontier Physics Research Division, Department of Physics and Astronomy, Seoul National University, Seoul, South Korea'' for financial support.
This research was supported by Basic Science Research Program through the National Research Foundation of Korea(NRF) funded by the Ministry of Education (2013057640).}\\
Lattice Gauge Theory Research Center, FPRD, and CTP \\
Department of Physics and Astronomy, Seoul National University,Seoul, 151-747, South Korea}

\abstract{We study the deconfinement transition in two-flavour lattice QCD with dynamical overlap fermions. Our simulations have been carried out on a $16^3 \times 6$ lattice at a pion mass around $500 \mev$ with a special HMC algorithm without any approximation such as fixed topology. We consider several temperatures from $220\mev$ which is close to the deconfinement to $280\mev$ which is above it. The dependence of the Polyakov loop, the chiral condensate, the Dirac spectra and the connected part of chiral susceptibility on the inverse gauge coupling has been studied. Our data indicates that the transition point lies between $\beta = 7.6$ and $\beta = 8.1$, but a more precise determination is not possible with our present statistics.}

\FullConference{The 32nd International Symposium on Lattice Field Theory,\\
23-28 June, 2014\\
Columbia University New York, NY}

\begin{document}

\section{Introduction and numerical setup}
\label{sec:intro}

 The study of the QCD thermodynamics, and in particular the confinement - deconfinement phase transition is one of the main applications of lattice QCD. Since the parameters of the deconfinement transition are of utmost importance for the interpretation of experimental data from heavy-ion colliders, one should reduce any systematical errors when studying them numerically. For this reason lattice QCD simulations with chirally invariant overlap fermions are now the state-of-the-art \cite{Fodor:04,Fodor:12,Cossu:13}. Unfortunately, these simulations are very expensive computationally and are almost always restricted to a single topological sector. Several important improvements in the Hybrid Monte-Carlo algorithm \cite{Arnold:2003sx, Cundy:2004pza, Cundy:06, Cundy:09:1, Cundy:09:2, Cundy:11:2} have allowed for large-scale simulations with dynamical overlap fermions and without any restriction to the fixed topology sector. While in general one cannot expect that the use of dynamical overlap fermions will result in a strong modification of thermodynamic properties, they can be very useful, e.g. to study fluctuations of topology at finite temperature.

Another important application of the algorithms of \cite{Arnold:2003sx, Cundy:2004pza, Cundy:06, Cundy:09:1, Cundy:09:2, Cundy:11:2} is the study of the deconfinement phase transition in an external magnetic field, which has attracted a lot of attention recently and was intensively investigated both theoretically \cite{Agasian:08:1,Fraga:08,Simonov:14} and in lattice simulations \cite{D'Elia:10,Endrodi:12:jhep,Muller-Preussker:13}. Lattice simulations \cite{D'Elia:10,Endrodi:12:jhep} have revealed a strong dependence of the sign of the shift of the deconfinement temperature in external magnetic field on the pion mass. In \cite{D'Elia:10} it was found that at sufficiently large pion masses the magnetic field increases the chiral condensate by the conventional ``magnetic catalysis'' mechanism \cite{Gusynin:94:1,Smilga:97:1} and hence increases the deconfinement temperature. On the other hand, the simulations of \cite{Endrodi:12:jhep} were performed at physical pion mass and revealed an unexpected decrease of the chiral condensate with magnetic field in the vicinity of the deconfinement transition, which results in a decrease of deconfinement temperature in magnetic field (``inverse magnetic catalysis''). At the same time, in recent theoretical works \cite{Chao:13,Yu:14:1} it was suggested that the fluctuations of chirality and topology can play an important role in the inverse magnetic catalysis. Thus the use of chiral lattice fermions with unrestricted topology can be advantageous for numerical studies of the inverse magnetic catalysis. Indeed, in our recent work \cite{Kochetkov:14} it was found that in HMC simulations with dynamical overlap fermions \cite{Arnold:2003sx, Cundy:2004pza, Cundy:06, Cundy:09:1, Cundy:09:2, Cundy:11:2} inverse magnetic catalysis is observed for pion masses as large as $500 \, {\rm MeV}$.

Unfortunately, up to now the exact location of the phase transition for the lattice action used in our dynamical overlap simulations \cite{Kochetkov:14} is not known. Since the knowledge of the critical temperature is an essential prerequisite for any further finite-temperature simulations with the algorithms of \cite{Arnold:2003sx, Cundy:2004pza, Cundy:06, Cundy:09:1, Cundy:09:2, Cundy:11:2}, in these Proceedings we report on our preliminary studies of the deconfinement phase transition for dynamical overlap fermions. We are able to identify the temperatures which certainty correspond to the confinement and the deconfinement regimes, however, the precise location of the phase transition remains elusive with our present statistics.

 We consider lattice QCD with $N_f = 2$ flavours of dynamical overlap fermions with equal masses. We use the massive overlap Dirac operator,
\begin{eqnarray}
\label{overlap_definition}
D\lrs{\mu} = 1 + \mu/2 + \gamma_5 \lr{1 - \mu/2} \sign\lr{K} ,
\end{eqnarray}
where $K = \gamma_5 \lr{D_W - \rho}$ and $D_W$ is the Wilson-Dirac operator with one level of over-improved stout smearing \cite{Moran:2008ra,Morningstar:2003gk}. In order to ensure that lattice gauge fields are sufficiently smooth, we use the tadpole improved L\"{u}scher-Weisz gauge action \cite{TILW, TILW2}. The temperature $T = 1/(N_t a)$ is changed by varying the inverse gauge coupling $\beta$ and thus the lattice spacing $a$.
The pion mass and lattice spacing were determined using independent runs on $12^3\times 24$ lattices for $\beta=7.5$ and $\beta=8.3$. We have performed measurements at $\beta=7.5$, which corresponds to $a=0.15\fm$ and $T=220\mev$ and at $\beta=8.3$, for which $a=0.12\fm$ and $T=280\mev$ as well as at the intermediate values of $\beta = 7.6, 7.7, 7.8, 7.9$ and $8.1$. For these intermediate values of $\beta$ the scale setting has not been performed yet. For every value of $\beta$ we have between $500$ and $1000$ successively generated configurations. The correlations between the configurations are taken into account using the Jackknife method.

\section{Numerical results}
\label{sec:num_res}

\begin{figure}[h!tpb!]
\includegraphics[angle=-90,width=7.7cm]{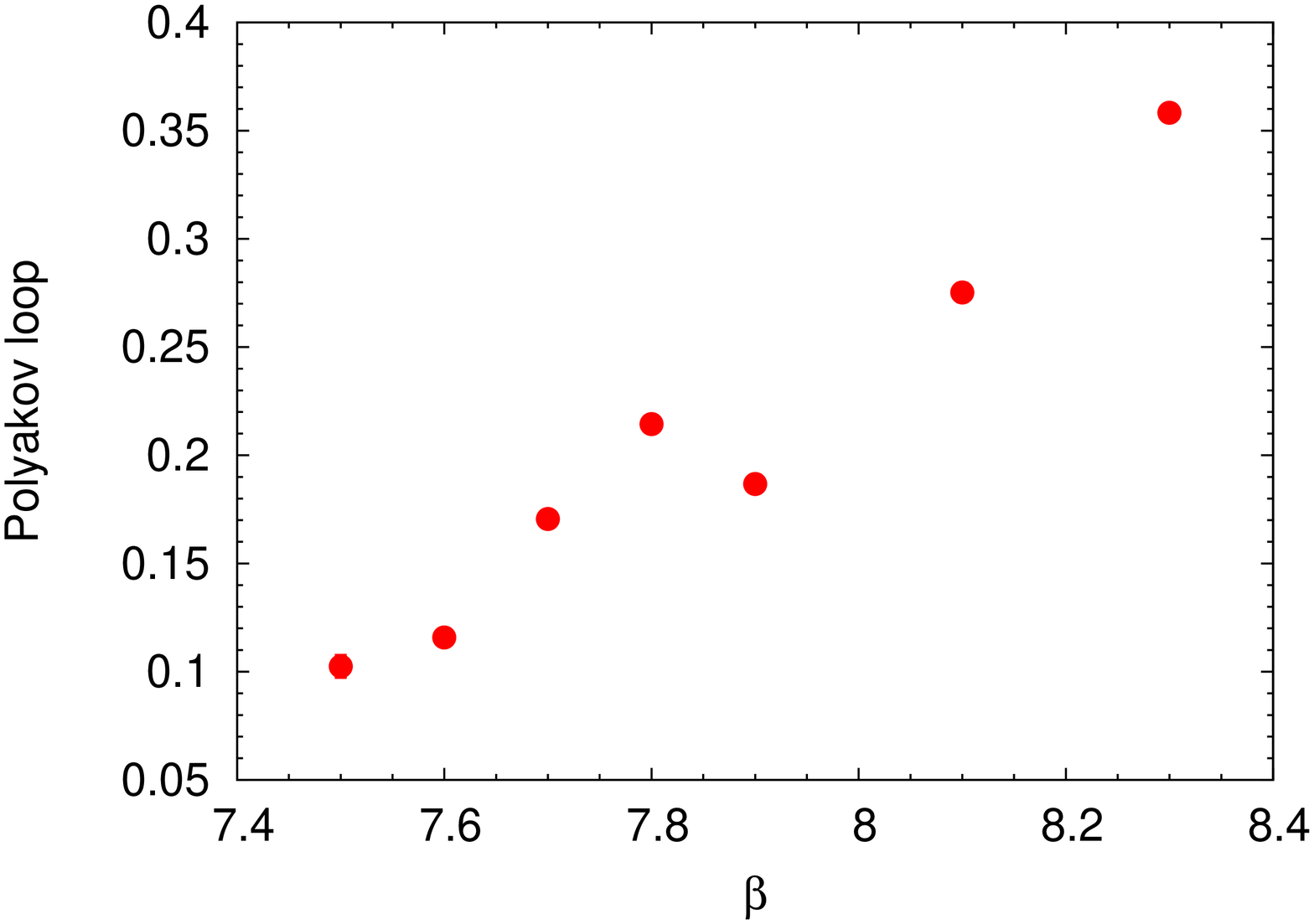}
\includegraphics[angle=-90,width=7.7cm]{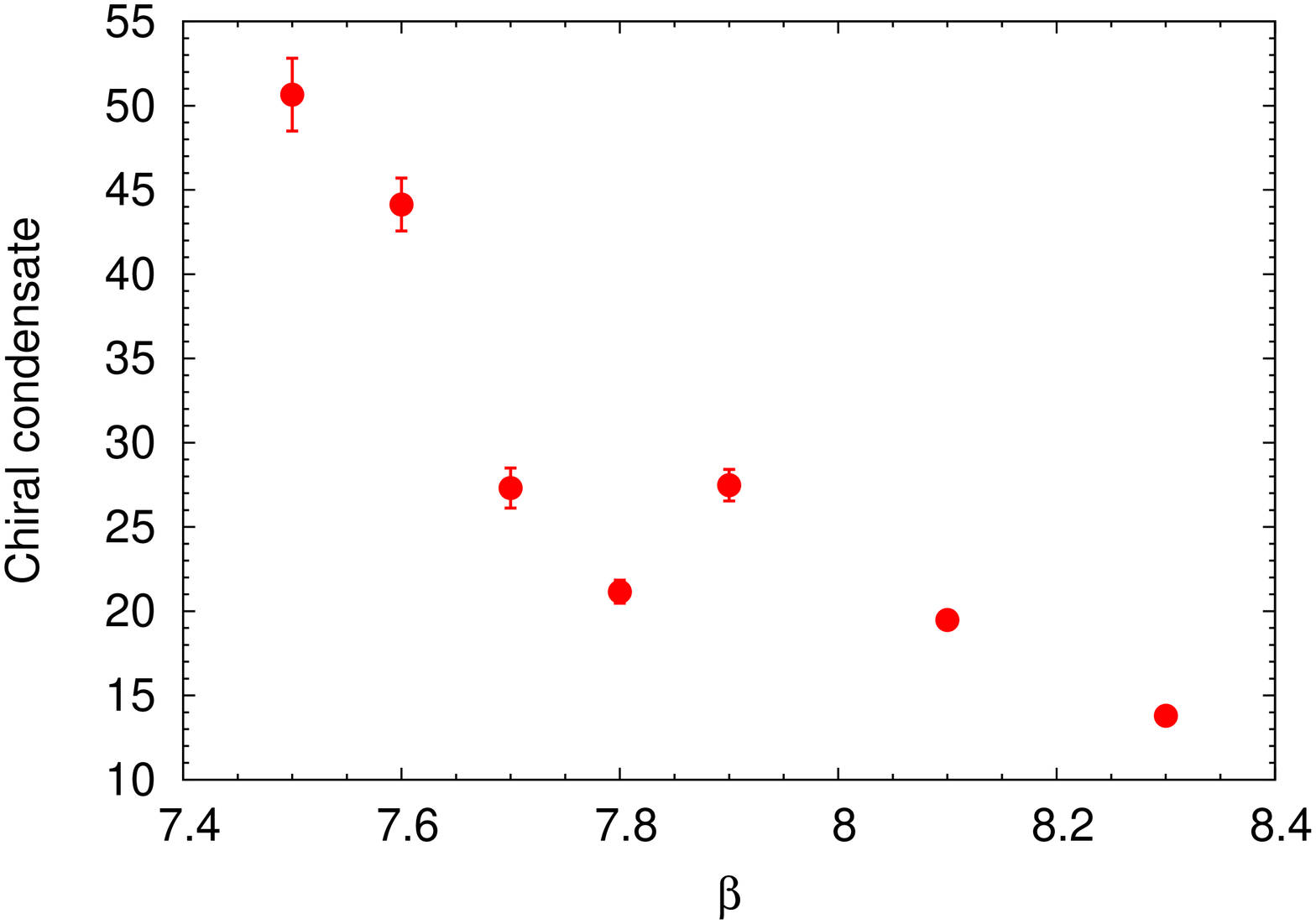}
\caption{Polyakov loop (left) and chiral condensate (right) as functions of the inverse gauge coupling $\beta$.}
\label{fig:pl_condensate}
\end{figure}

\textit{Polyakov loop.} \quad The expectation values of the Polyakov loop are shown in Fig.~\ref{fig:pl_condensate} at different inverse gauge couplings. As expected, the Polyakov loop increases at larger temperatures. Unfortunately, from our data it is difficult to identify the inflection point of the Polyakov loop. Also, the points at $\beta = 7.7$ and $7.8$ deviate somehow from the smooth behavior, which might be the result of some long-range correlation in our simulations.

\textit{Chiral condensate.} The chiral condensate in lattice units is shown in Fig.~\ref{fig:pl_condensate} on the right. Again, the condensate gradually decreases towards larger inverse gauge couplings, except for some deviation at points $\beta = 7.7$ and $\beta = 7.8$. There is also no clearly defined inflection point.

\begin{figure}[h!tpb!]
\centering
\includegraphics[angle=-90,width=7.7cm]{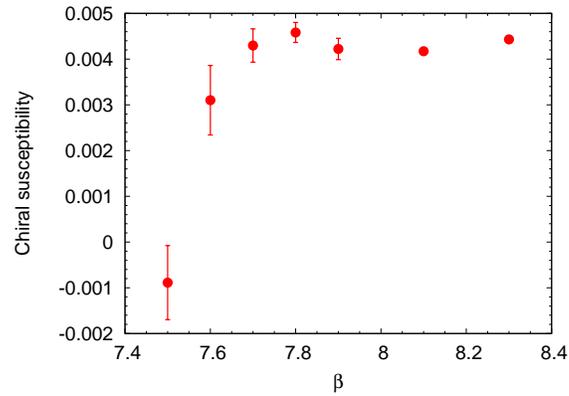}
\caption{Connected part of the chiral susceptibility (3.1) as a function of the inverse gauge coupling $\beta$.}
\label{fig:susc_vev}
\end{figure}

\textit{Chiral susceptibility.} A quantity which is typically more sensitive to the (partial) restoration of chiral symmetry than the expectation value of the chiral condensate is the chiral susceptibility
\begin{equation}
\label{susc_def}
\chi_c(T)= -\left.
\frac{\partial \langle {\bar u}u\rangle}{\partial m_q} \right|_{m_q=0} .
\end{equation}
At the transition point, $\chi_c\lr{T}$ usually has a characteristic peak with height which increases with volume for the first or second order phase transitions and stays constant for the crossover. In Fig.~\ref{fig:susc_vev} we show the connected part of the chiral susceptibility (in lattice units) as a function of $\beta$. Unfortunately, we do not see a pronounced peak, but rather some sort of plateau starting from $\beta = 7.6$. There is only a slight hint at the peak at $\beta = 7.8$.

\begin{figure}[h!tpb!]
\includegraphics[angle=0,width=14cm]{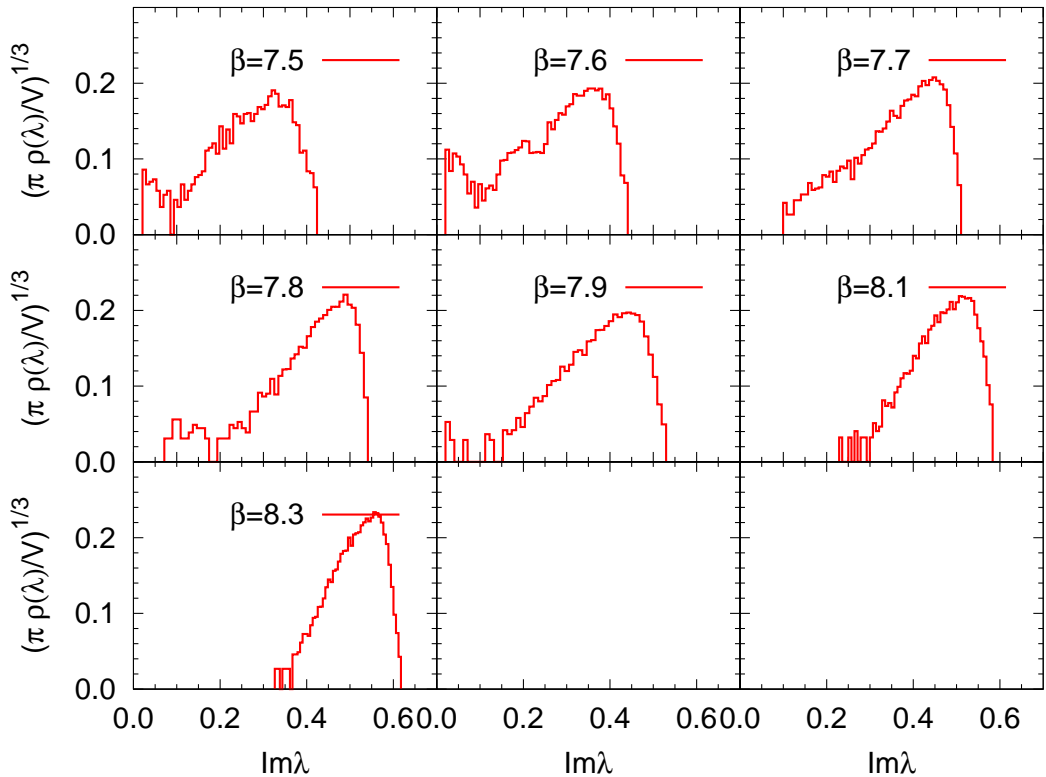}
\caption{Histograms of the eigenvalues $\lambda$ of the massless overlap Dirac operator in lattice units at different values of the inverse coupling constant $\beta$ which correspond to different temperatures in the range $220 \mev < T < 280 \mev$.}
\label{fig:hist}
\end{figure}

\textit{Low-lying Dirac eigenvalues.} As a more sensitive test of the temperature at which the chiral symmetry is restored, we consider the statistical distributions of the low-lying eigenvalues $\lambda$ of the projected massless Dirac operator
\begin{eqnarray}
\label{massless_dirac}
\tilde{D}_0 = \frac{2 \rho D_0}{2 - D_0}, \quad D_0 = 1 + \gamma_5 \sign\lr{K} .
\end{eqnarray}
The eigenvalues $\lambda$ of $\tilde{D}_0$ are purely imaginary and are related to the chiral condensate on the lattice exactly in the same way as in the continuum theory:
\begin{eqnarray}
\label{lattice_condensate}
\Sigma = \sum\limits_i \frac{1}{m_q + \lambda_i} = \sum\limits_{\Im \lambda_i > 0} \frac{2 m_q}{m_q^2 + |\lambda_i|^2},
\end{eqnarray}
where $\Sigma = \frac{1}{V} \frac{\partial}{\partial \, m_q} \mathcal{Z}\lr{m_q}$ and $\mathcal{Z}\lr{m_q}$ is the lattice partition function with the Dirac operator (\ref{overlap_definition}). By virtue of the relation (\ref{lattice_condensate}), which implies that the condensate is mostly saturated by Dirac eigenmodes with $|\lambda_i| \lesssim m_q$, effective restoration of chiral symmetry should result in a significant widening of the gap in the spectrum of $\tilde{D}_0$.

The histograms of $\Im \lambda$ (plotted in lattice units) are shown in Figure~\ref{fig:hist}. At $\beta=7.5$ and $\beta=7.6$ one can see a lot of near-zero eigenvalues which indicates that these two points are still in the phase with broken chiral symmetry. For $\beta = 7.7$ and $\beta = 7.8$, it seems that the gap already starts to appear. However, at $\beta = 7.9$ some near-zero eigenvalues appear again. This non-monotonic behavior of the gap agrees with the non-monotonic behavior of the Polyakov loop and the chiral condensate at $\beta = 7.7$ and $\beta = 7.8$. Finally, at $\beta=8.1$ and $\beta=8.3$ the gap in the spectrum of $\lambda$ is well pronounced, as expected for the phase with restored chiral symmetry.

\begin{figure}[h!tpb!]
\includegraphics[angle=-90,width=7.7cm]{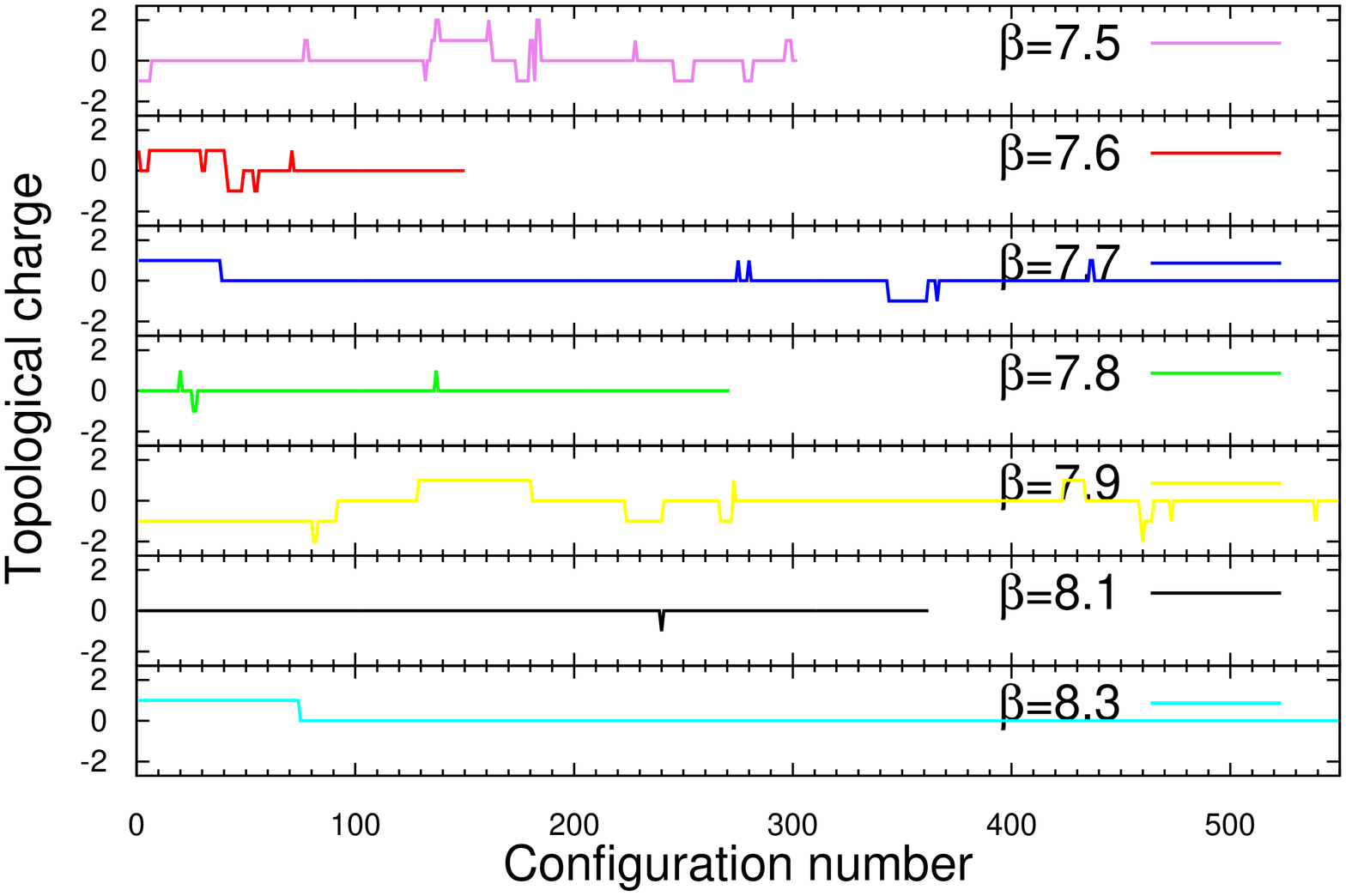}
\includegraphics[angle=-90,width=7.7cm]{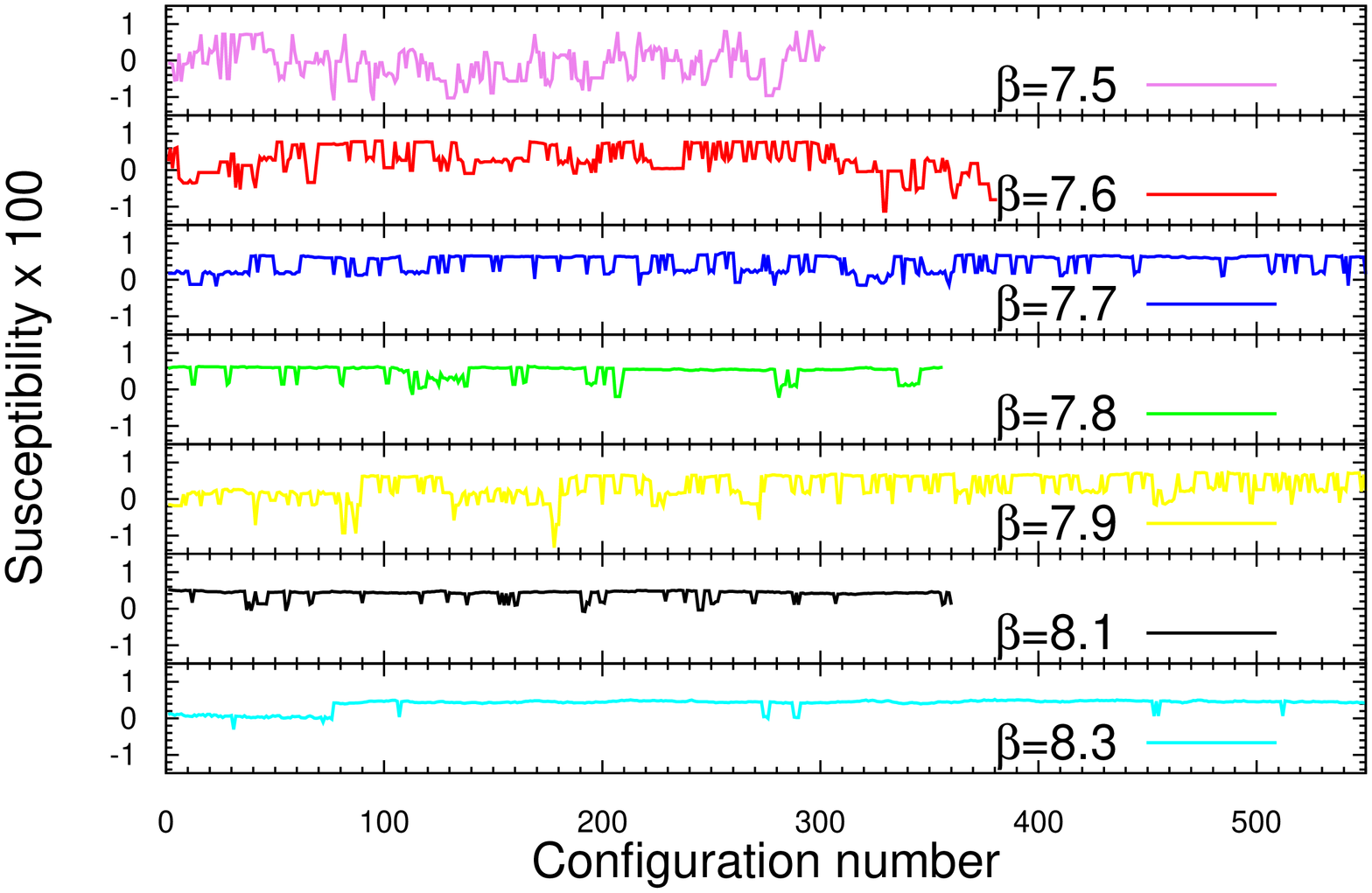}
\caption{Monte-Carlo histories of of the topological charge (left) and the connected part of the chiral susceptibility (right).}
\label{fig:mc_histories}
\end{figure}

\textit{Topological charge fluctuations.} Yet another possible way to distinguish the confinement and the deconfinement phases is to consider the fluctuations of topological charge, which should be strongly suppressed above the deconfinement temperature \cite{Shuryak:98:1}. Monte-Carlo histories of the topological charge (defined from the number of exact zero modes of the massless overlap Dirac operator (\ref{massless_dirac})) are shown in Figure~\ref{fig:mc_histories}(left) for different values of $\beta$. There are significant fluctuations of topological charge at $\beta=7.5,7.6$ and very few fluctuations at $\beta = 7.7, \, 7.8, \, 8.1, \, 8.3$. Strangely, at $\beta = 7.9$ strong fluctuations reappear. Again we see that the points with $\beta = 7.7$ and $\beta = 7.8$ somehow deviate from the general trend.

The autocorrelation time of topological charge in our simulations is of the order of several hundreds of HMC trajectories. Of course, this is only a rough estimate based on our sets of no more than thousand configurations. Let us also note that long autocorrelations of the topological charge might significantly affect the autocorrelation time of the chiral susceptibility. A close look at the Monte-Carlo histories of the connected chiral susceptibility (see Fig.~\ref{fig:mc_histories} on the right) reveals that it is strongly correlated with the topological charge. When the topological charge changes, the value of the susceptibility also shifts to some ``plateau'' and then fluctuates around this ``plateau'' (see e.g. the HMC histories for $\beta = 7.9$ near configurations $100$ and $200$). The resulting large autocorrelations of the chiral susceptibility might explain the absence of well-defined peak in the chiral susceptibility in Fig.~\ref{fig:susc_vev}.

\section{Discussion and conclusions.}
\label{sec:conclusions}

The preliminary data presented here certainly rules out the phase transition (or crossover) at $\beta \leq 7.6$ and $\beta \geq 8.1$, thus the transition point in our simulations with dynamical overlap fermions should be somewhere in the range $7.6 < \beta < 8.1$. Within this range of $\beta$, however, the precision of our measurements is insufficient to determine the transition point. The fact that for $\beta = 7.7$ and $\beta = 7.8$ all our observables deviate from the smooth behavior suggests that autocorrelation times might be larger for these two points, which can in turn hint at the proximity to the phase transition. We therefore plan to investigate this region of $\beta$ values using larger numbers of configurations. We should also note that at present we perform simulations at fixed bare quark mass, thus changing the temperature by varying the inverse gauge coupling $\beta$ results also in some variation of the pion mass. Therefore in our simulations the physical pion mass is not fixed. Possible improvement of our simulation strategy would be then either to keep the physical pion mass constant, or to vary the temperature by changing the temporal size of the lattice.

%\bibliographystyle{mybibstyle}
%\bibliography{Kochetkov}

\end{document}